# Evaluation of Quantum Offset in Velocity Imaging-Based Electron Spectrometry

Rui Zhang, Shuaiting Yan, Wenru Jie, Jiayi Chen, Qihan Liu, and Chuangang Ning *

*Department of Physics, State Key Laboratory of Low Dimensional Quantum Physics,
Frontier Science Center for Quantum Information, Tsinghua University, Beijing 100084, China*

Velocity-map imaging of electrons is a pivotal technique in chemical physics. A recent study reported a quantum offset as large as 0.2 cm$^{-1}$ in velocity imaging-based electron spectrometry [Phys. Rev. Lett. **134**, 043001 (2025)]. In this work, we assess the existence this offset through a combination of simulations and experiments. Our simulations reveal that the velocity imaging results reconstructed using the maximum entropy velocity Legendre reconstruction (MEVELER) method exhibit no such offset. Furthermore, experimental measurements of the electron affinity of oxygen conducted at various imaging voltages show no discernible offset attributable to the electric field in the photodetachment region. Therefore, we conclude that there is no evidence for the claimed quantum offset in properly analyzed velocity imaging-based electron spectrometry.

*Introduction*—Velocity-map imaging (VMI) has revolutionized the measurement of charged-particle energies by projecting electrons or ions onto a two-dimensional detector, thereby directly converting velocity information into spatial coordinates of the detection plane [1]. This projection-based approach was later adapted for negative ion photoelectron spectroscopy, establishing VMI as a key method for simultaneously probing electron energy and angular distributions [2,3]. Recently, a specifically designed VMI variant for low-energy electrons, known as slow electron velocity-map Imaging (SEVI) method, has achieved an impressive energy resolution of ~1 cm$^{-1}$ [4-9]. This high resolution makes SEVI particularly valuable for obtaining precise measurements of fundamental properties such as electron affinities [10,11].

In a recent study, Blondel and Drag claimed that there is a quantum offset as large as ~ 0.2 cm$^{-1}$ in photoelectron kinetic energy measurements using velocity imaging-based spectrometry [12]. They attribute this offset to the difference between the maximum classical trajectory radius $R_c$ and the electron current maximum radius $R_0$ on the 2D detector in the laser photodetachment microscopy (LPM) [13]. From a semi-classical perspective, a detached photoelectron with low kinetic energy $E_k$ follows a parabolic trajectory in the uniform imaging electric field $F$. For a given point on the detector that the electron can reach, there exist two possible trajectories with a phase difference, leading to interference that forms a characteristic pattern. The maximum classical trajectory radius $R_c$ falling off the axis, at which electron trajectories become degenerate, can be calculated as $R_c \approx 2\sqrt{zE_k/qF}$ with high accuracy, where $q$ is the charge of the photoelectron and $z$ is the distance between the detachment point and the detector [12,13]. There is a relationship between these two radii: $R_0^2 = R_c^2 + 4zx_0\lambda_0$, where $x_0 = -1.01879...$ is the first maximum of the Airy function, and $\lambda_0$ is the characteristic de Broglie wavelength of the electron in the acceleration field [12]. They further proposed that the energy offset Δ (in cm$^{-1}$), caused by the gap between $R_0$ and $R_c$, depends solely on the electric field $F$ (in V/m) in the photodetachment region as $\Delta = 0.002765..F^{2/3}$ [12]. However, their analysis completely neglected the crucial aspect of SEVI: the reconstruction process implemented via the maximum entropy velocity Legendre reconstruction (MEVELER) method [14]. In SEVI, the kinetic energy $E_k$ is derived from the weighted center radius $R_s$ of the photoelectron velocity distribution, obtained by summing the reconstructed three-dimensional spherical shell over all angles, not from $R_0$ as claimed in their work [11,15].

The assertion of a substantial quantum offset has raised considerable concern in the field, as it challenges the foundational accuracy of velocity-map imaging. To address this discrepancy, a rigorous evaluation of the alleged quantum offset is essential. In this Letter, we present a combined simulation and experimental study to assess the alleged quantum offsets in velocity imaging-based spectrometry.

*Simulation of photodetachment imaging*—An ideal photodetachment process was simulated under a uniform imaging electric field of $F$ = 544 V/m, corresponding to an imaging voltage $V$ = 75 V in our spectrometer, as illustrated in Fig. 1. The detachment laser energy was set to exceed the binding energy of a hypothetical photodetachment transition by 5 cm$^{-1}$, with the detachment point located at a distance $z$ = 0.5 m from the 2D detector. Theoretically, due to their wave nature, the detached photoelectrons form an Airy-disk-like intensity distribution at the detector plane [12] (Fig. 1a). To accurately mimic our measurement, this ideal distribution was numerically convolved with a 2D Gaussian function, ($Ae^{\frac{-4ln2\cdot(x^2+y^2)}{w^2}}$), to account for the instrumental response of the VMI spectrometer (Fig. 1b). Here $A$ is the normalization coefficient and $w$ = 2.8 pixels is the full width at half maximum (FWHM) of the Gaussian profile. The resulting signal was then discretized by a pixelated detector (1024 × 1024 pixels). In typical analysis, the acquired photoelectron image is reconstructed either by the MEVELER method or an inverse Abel transform [16] to retrieve the underlying 3D velocity distribution. A central cross-section of the reconstructed distribution is shown in Fig.1c. The corresponding radial distributions for Fig.1 a-c are plotted in Fig.1 d-f, where three characteristic radii are indicated: $R_0$ (the radius of maximum intensity on the detector), $R_c$ (the classical maximum radius), and $R_s$ (the weighted center radius obtained from Gaussian fitting of the reconstructed distribution). Notably, the reconstructed radius $R_s$ is nearly identical to the classical radius $R_c$, in clear contrast to the assumption in Ref. [12], which relied on $R_0$.

*Contact author: ningcg@tsinghua.edu.cn

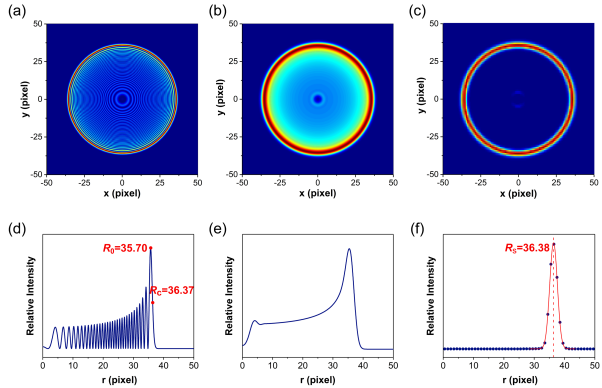

FIG. 1. Simulation of photodetachment imaging. The simulation was performed under an electric field of $F$ = 544 V/m (imaging voltage $V_i$ = 75 V), with the detachment region located at $z$ = 0.5 m from the detector. The photon energy was set to 5 cm$^{-1}$ above the binding energy of the simulated transition. (a) Theoretical two-dimensional distribution of photoelectrons at the detector plane. (b) Convolved photoelectron image after accounting for the Gaussian response of the spectrometer. (c) Image reconstructed using the MEVELER method. (d)–(f) Radial distributions corresponding to (a)–(c), respectively.

To further quantitatively assess the measurement offset associated with the three radii ($R_0$, $R_c$, and $R_s$), we systematically varied the detachment laser energy $hv$ in simulation from 3 to 8 cm$^{-1}$ above the photodetachment threshold and recorded the corresponding radii, as summarized in Table I. The squared radii were then extrapolated as a function of the energy difference ($hv$ − BE), with the intercept of the linear fit indicating the binding energy offset for each radius (Fig. 2). Using the classical maximum radius $R_c$ as a benchmark yields a zero offset, as expected. Importantly, the radius $R_s$ obtained via the MEVELER method also gives a negligible offset 0.004(2) cm$^{-1}$, where the uncertainty of 0.002 cm$^{-1}$ represents one standard deviation from the fitting procedure. In contrast, extrapolation using $R_0$ results in a significant offset of 0.184 cm$^{-1}$, consistent with the prediction in Ref. [12]. These simulation results show that the quantum offset arising from the difference between $R_s$ and $R_c$ is negligible. Hence, no quantum offset is present in properly analyzed velocity imaging-based electron spectrometry.

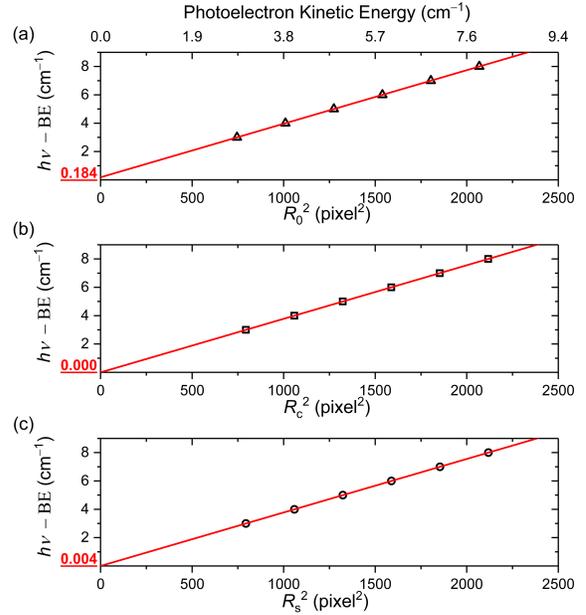

FIG. 2. Squared radii of the photoelectron spherical shells versus the energy difference ($hv$ − BE) for the simulated transition at imaging voltage $V_i$ = 75 V. The binding energy offset for each radius, given by the intercept of a linear least-squares fit, is shown for: (a) $R_0^2$, (b) $R_c^2$, and (c) $R_s^2$.

TABLE I. Simulated values of the three characteristic radii ($R_0$, $R_c$, and $R_s$) at various detachment laser photon energy. Energies and radii are given in units of cm$^{-1}$ and pixels, respectively. The uncertainty of $R_s$ represents one standard deviation from the fitting procedure.

| $hv$ − BE | $R_0$ | $R_c$ | $R_s$ |
|---|---|---|---|
| 3 | 27.296 | 28.175 | 28.173(11) |
| 4 | 31.776 | 32.534 | 32.535(8) |
| 5 | 35.698 | 36.374 | 36.380(7) |
| 6 | 39.230 | 39.846 | 39.858(7) |
| 7 | 42.468 | 43.039 | 43.056(5) |
| 8 | 45.477 | 46.010 | 46.020(5) |

*Experimental offset assessments*— It is important to note that, unlike uniform electric fields employed in laser photodetachment microscopy (LPM) [12,13], the electric fields in typical velocity-map imaging apparatus are not uniform [1,5]. In our SEVI setup, a deliberately non-uniform electric field is applied within the imaging zone (0.1 m) to focus photoelectrons—originating from different initial positions but possessing identical velocities—onto a common point on the detector. After passing through the imaging lens, photoelectrons travel through a field-free zone (0.4 m) before hitting the detector. The electric field in the photodetachment region corresponding to each experimental imaging voltage was determined via numerically simulation of our VMI lens system.

To experimentally evaluate the potential quantum offset in the SEVI method, we measured the electron affinity (EA) of oxygen at various imaging voltages. Figure 3 plots the deviation of our measured EA values from the reference EA(O) value of 11784.671(1) cm$^{-1}$ [17] as a function of imaging voltage $V_i$. Each EA value has been extracted using the procedure illustrated in Fig.1f and Fig.2c. The dotted curve in Fig.3 represents the offsets calculated using this formula $\Delta = 0.002765..F^{2/3}$, using the electric field specific to the photodetachment zone. Since the electric field in our VMI spectrometer is non-uniform, the offset calculated using the average field over the entire 0.5 m flight path is also shown for comparison (dashed curve). As clearly shown in Fig.3, our results show no discernible offset from the highly accurate reference value. This stands in sharp contrast to the substantial deviations predicted in Ref. [12].

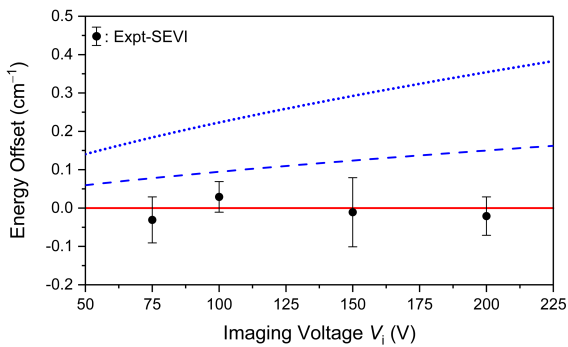

FIG. 3. Experimental test of the quantum offset claim via SEVI measurements of the oxygen electron affinity. The measured deviation from the reference value [17] is plotted against the imaging voltage. The dotted curve is the calculated offset $\Delta$ using the field in the photodetachment zone, while the dashed curve using the average field over the whole flight path ($F = V_i/0.5$ m).

To further validate the absence of the claimed quantum offset across multiple systems, Table II summarizes EA values of several main-group elements measured by SEVI and other methods. Among these, SEVI results of EA(O) and EA(Sb) were obtained in the current work. EA(O) is the average measurement results at the four voltages shown in Figure 3. Other SEVI data were acquired at an imaging voltage of 150 V, which, according to Ref. [12], would have a predicted offset of 0.292 cm$^{-1}$. The differences between SEVI and other methods show no consistent positive deviation, contrary to the prediction of Ref. [12], indicating that the purported quantum offset is absent in SEVI measurements. The previous EA(Sb) value, measured using the laser photodetachment threshold (LPT) method, carries a relatively large uncertainty of 0.15 cm$^{-1}$ [18], and an independent, more accurate measurement is therefore needed for a definitive comparison with our EA(Sb) = 8447.70(6) cm$^{-1}$. Furthermore, we recently found that the LPM-measured EA(Se) exhibits a significant deviation of 0.5 cm$^{-1}$ [19], indicating that LPM alone may not serve as a reliable benchmark for validating other methods.

TABLE II. Electron affinities of several main-group elements: SEVI vs other methods. The SEVI-Other difference is shown in the last column. All values are in cm$^{-1}$.

| $_z$Element | SEVI | Other | Difference |
|---|---|---|---|
| $_8$O | 11784.66(4)[a] | 11784.671(1)[b] | -0.011(41) |
| $_{16}$S | 16753.00(7)[c] | 16752.9753(41)[d] | +0.025(74) |
| $_{33}$As | 6488.61(5)[c] | 6488.570(15)[e] | +0.04(6) |
| $_{51}$Sb | 8447.70(6)[a] | 8447.86(15)[f] | -0.16(17) |
| $_{82}$Pb | 2877.14(9)[g] | 2877.149(15)[h] | -0.009(92) |

[a] SEVI (present work); [b] LPT [17]; [c] SEVI [11]; [d] LPM [21]; [e] LPT [12]; [f] LPT [18]; [g] SEVI [22]; [h] LPM [23].

*Discussion and Conclusion*—Combined evidence from our simulations and experiments demonstrates conclusively that no quantum offset exists in SEVI spectrometry when the data are analyzed using an appropriate reconstruction methodology. The central point of disagreement lies in the identification of the relevant measured quantity in velocity-map imaging. The hypothesis proposed by Blondel and Drag [12] rests on the assumption that the key radius in velocity imaging-based electron spectrometry is the point of maximum intensity on the raw detector image, ($R_0$). In contrast, the standard SEVI analysis correctly extracts the expectation value of the radius, $R_s$, from the reconstructed 3D velocity distribution, which aligns with the classical radius $R_c$. In real SEVI measurements, the VMI lens system projecting the 3D photoelectron velocity distribution onto a 2D detector, may introduce distortions due to the nonperfect effects of the imaging lens and 2D detector, and the sizes of laser beam and anion beam. These effects were taken into account via the instrumental response function in the simulation. Since the system has cylindric symmetry, the original 3D photoelectron velocity distribution can be reconstructed from its 2D projection. The resulting radius $R_s$ corresponds to the weighted center of the initial velocity shell, consistent with quantum mechanics, represents the expectation value of the radius. Therefore, the comparison made in Ref. [12] between a classical trajectory radius $R_c$ and a raw image feature $R_0$ is not appropriate for evaluating a properly analyzed SEVI experiment. The correct counterpart to $R_c$ is the reconstructed radius $R_s$. This correspondence is further supported by the Ehrenfest theorem, which states that the quantum-mechanical expectation values follow classical equations of motion. For a particle moving in a potential $U$, the time derivative of its average momentum $\langle \dot{p} \rangle$ satisfies $\langle \dot{p} \rangle = -\nabla U(z)$, a relation that applies in both quantum and classical regimes [20]. As a result, no quantum offset arises between $R_c$ and $R_s$.

In summary, through simulations and experiments, we have shown that no universal quantum offset exists in velocity imaging-based electron spectrometry when properly data analysis is employed, and specifically, that no such offset is present in SEVI. Our findings affirm that the SEVI method, as widely practiced, is fundamentally sound and highly reliable for obtaining precise spectroscopic quantities such as electron affinities.

*Acknowledgments*—This work was supported by the National Natural Science Foundation of China (NSFC) (Grant Nos. 12374244 and 12341401).

[1] A. T. J. B. Eppink and D. H. Parker, *Velocity map imaging of ions and electrons using electrostatic lenses: application in photoelectron and photofragment ion imaging of molecular oxygen*, Rev. Sci. Instrum. **68**, 3477 (1997).

[2] E. Surber and A. Sanov, *Photoelectron imaging spectroscopy of molecular and cluster anions: $CS_2^-$ and $OCS^-(H_2O)_{1,2}$*, J. Chem. Phys. **116**, 5921 (2002).

[3] A. Osterwalder, M. J. Nee, J. Zhou, and D. M. Neumark, *High resolution photodetachment spectroscopy of negative ions via slow photoelectron imaging*, J. Chem. Phys. **121**, 6317 (2004).

[4] S. Cavanagh, S. Gibson, M. Gale, C. J. Dedman, E. Roberts, and B. Lewis, *High-resolution velocity-map-imaging photoelectron spectroscopy of the $O^-$ photodetachment fine-structure transitions*, Phys. Rev. A **76**, 052708 (2007).

[5] I. León, Z. Yang, H.-T. Liu, and L.-S. Wang, *The design and construction of a high-resolution velocity-map imaging apparatus for photoelectron spectroscopy studies of size-selected clusters*, Rev. Sci. Instrum. **85** (2014).

[6] D. M. Neumark, *Slow electron velocity-map imaging of negative ions: Applications to spectroscopy and dynamics*, J. Phys. Chem. A **112**, 13287 (2008).

[7] C. Hock, J. B. Kim, M. L. Weichman, T. I. Yacovitch, and D. M. Neumark, *Slow photoelectron velocity-map imaging spectroscopy of cold negative ions*, J. Chem. Phys. **137** (2012).

[8] M. L. Weichman and D. M. Neumark, *Slow photoelectron velocity-map imaging of cryogenically cooled anions*, Annu. Rev. Phys. Chem. **69**, 101 (2018).

[9] R. Tang, X. Fu, Y. Lu, and C. Ning, *Accurate electron affinity of Ga and fine structures of its anions*, J. Chem. Phys. **152** (2020).

[10] C. Ning and Y. Lu, *Electron Affinities of Atoms and Structures of Atomic Negative Ions*, J. Phys. Chem. Ref. Data **51** (2022).

[11] S. Yan, Y. Lu, R. Zhang, and C. Ning, *Electron affinities in the periodic table and an example for As*, Chin. J. Chem. Phys. **37**, 1 (2024).

[12] C. Blondel and C. Drag, *Quantum Offset of Velocity Imaging-Based Electron Spectrometry and the Electron Affinity of Arsenic*, Phys. Rev. Lett. **134**, 043001 (2025).

[13] C. Blondel, C. Delsart, and F. Dulieu, *The photodetachment microscope*, Phys. Rev. Lett. **77**, 3755 (1996).

[14] B. Dick, *MELEXIR: maximum entropy Legendre expanded image reconstruction. a fast and efficient method for the analysis of velocity map imaging or photoelectron imaging data*, Phys. Chem. Chem. Phys. **21**, 19499 (2019).

[15] D. M. Neumark, *Spectroscopy of radicals, clusters, and transition states using slow electron velocity-map imaging of cryogenically cooled anions*, J. Phys. Chem. A **127**, 4207 (2023).

[16] V. Dribinski, A. Ossadtchi, V. A. Mandelshtam, and H. Reisler, *Reconstruction of Abel-transformable images: The Gaussian basis-set expansion Abel transform method*, Rev. Sci. Instrum. **73**, 2634 (2002).

[17] M. K. Kristiansson *et al.*, *High-precision electron affinity of oxygen*, Nat. Commun. **13**, 5906 (2022).

[18] M. Scheer, H. K. Haugen, and D. R. Beck, *Single-and multiphoton infrared laser spectroscopy of $Sb^-$: A case study*, Phys. Rev. Lett. **79**, 4104 (1997).

[19] R. Zhang, W. Jie, J. Chen, Q. Liu, and C. Ning, *Revisiting the electron affinity of selenium*, arXiv preprint arXiv:2506.10300 (2025).

[20] J. J. Sakurai and J. Napolitano, *Modern quantum mechanics* (Cambridge University Press, 2020).

[21] T. Carette, C. Drag, O. Scharf, C. Blondel, C. Delsart, C. Froese Fischer, and M. Godefroid, *Isotope shift in the sulfur electron affinity: Observation and theory*, Phys. Rev. A **81**, 042522 (2010).

[22] C. X. Song *et al.*, *Isotope shifts in electron affinities and in binding energies of Pb and hyperfine structure of $^{207}Pb^-$*, J. Chem. Phys. **160** (2024).

[23] D. Bresteau, C. Drag, and C. Blondel, *Electron affinity of lead*, J. Phys. B: At. Mol. Opt. Phys. **52**, 065001 (2019).